\documentstyle[aps,preprint,eqsecnum,floats,epsf]{revtex}
\tighten
\draft
\begin{document}
 
\title{Algebraic Approach to Shape Invariance}
\author{A.~B. Balantekin\thanks{Electronic address:
        {\tt baha@nucth.physics.wisc.edu}}}
\address{Department of Physics, University of Wisconsin\\
         Madison, Wisconsin 53706 USA\thanks{Permanent Address}\\
         and\\
         Institute for Nuclear Theory, University of Washington, Box
         351550\\
         Seattle, WA 98195-1550 USA\\
         and\\
         Department of Astronomy, University of Washington, 
          Box 351580\\
       Seattle WA 98195-1580 USA}
\date{November 10, 1997}
\maketitle

\begin{abstract}
The integrability condition called shape invariance is shown to have 
an underlying algebraic structure and the associated Lie algebras are
identified. These shape-invariance algebras transform the parameters of the
potentials such as strength and range. Shape-invariance algebras, in
general, are shown to be infinite-dimensional. The conditions under
which they become finite-dimensional are explored.  
\end{abstract}



\newpage
\section{Introduction}

\indent

Supersymmetric quantum mechanics \cite{witten} and its connection to
the factorization method \cite{infeld} has been extensively
investigated \cite{coop1}. Since the ground state wavefunction,
$\psi_0(x)$, for a bound system has no nodes it can be written as 
\begin{equation}
  \label{1}
\psi_0(x) = \exp \left( - \frac{\sqrt{2m}}{\hbar} \int W(x) dx \right).
\end{equation}
Introducing the operators
\begin{eqnarray}
  \label{2}
  \hat{A} &=& W(x) + \frac{i}{\sqrt{2m}} \hat{p} \nonumber \\
  \hat{A}^{\dagger} &=& W(x) - \frac{i}{\sqrt{2m}} \hat{p}, 
\end{eqnarray}
the Hamiltonian can be easily factorized
\begin{equation}
  \label{3}
  \hat{H} - E_0 = \hat{A}^{\dagger}\hat{A}, 
\end{equation}
where $E_0$ is the ground state energy. Since the ground state
wavefunction satisfies the condition 
\begin{equation}
  \label{4}
  \hat{A}| \psi_0 \rangle = 0, 
\end{equation}
the supersymmetric partner potentials 
\begin{eqnarray}
  \label{5}
  \hat{H}_1 &=& \hat{A}^{\dagger} \hat{A} \nonumber \\
  \hat{H}_2 &=& \hat{A} \hat{A}^{\dagger}
\end{eqnarray}
have the same energy spectra except the ground state of $\hat{H}_1$
which has no corresponding state in the spectra of $\hat{H}_2$. The
corresponding potentials are given by 
\begin{eqnarray}
  \label{6}
  V_1(x) &=& [W(x)]^2 - \frac{\hbar}{\sqrt{2m}} \frac{dW}{dx} \nonumber
  \\
  V_2(x) &=& [W(x)]^2 + \frac{\hbar}{\sqrt{2m}} \frac{dW}{dx} . 
\end{eqnarray}

It was shown that a subset of the potentials for which the
Schr\"odinger equations are exactly solvable share an integrability
condition called shape invariance \cite{genden}. The partner
potentials of Eq. (\ref{6}) are called shape invariant if they satisfy
the condition 
\begin{equation}
  \label{7}
  V_2(x;a_1) = V_1 (x;a_2) + R(a_1),
\end{equation}
where $a_{1,2}$ are a set of parameters that specify space-independent
properties of the potentials (such as strength, range, diffuseness,
etc.), $a_2$ is a function of $a_1$, and the remainder $R(a_1)$ is
independent of $x$. One should emphasize that shape-invariance is not
the most general integrability condition as not all exactly solvable
potentials seem to be shape-invariant \cite{joe}. The purpose of 
this article is to show
that shape invariance has an underlying algebraic structure and 
to identify the associated Lie algebras. 


\section{Algebraic properties of shape invariance}

\indent

The shape invariance condition of Eq. (\ref{7}) can be rewritten in
terms of the operators defined in Eq. (\ref{2}) 
\begin{equation}
  \label{8}
  \hat{A}(a_1)\hat{A}^{\dagger}(a_1) = \hat{A}^{\dagger}(a_2)
  \hat{A}(a_2) + R(a_1), 
\end{equation}
where $a_2$ is a function of $a_1$. We assume that replacing $a_1$
with $a_2$ in a given operator can be achieved with a similarity
transformation: 
\begin{equation}
  \label{9}
  \hat{T}(a_1) {\cal O}(a_1) \hat{T}^{-1}(a_1) = {\cal O}(a_2).
\end{equation}
Such a transformation was used to construct coherent states for
shape-invariant potentials \cite{fukui}. So far two classes of
shape-invariant potentials are found. In the first class the
parameters $a_1$ and $a_2$ are related by a translation
\cite{joe,chuan}: 
\begin{equation}
  \label{10}
  a_2 = a_1 + \eta, 
\end{equation}
and in the second class they are related by a scaling \cite{khare}
\begin{equation}
  \label{11}
  a_2 = q a_1.
\end{equation}
All textbook examples of exactly solvable potentials belong to the
first class. In this article for definiteness we focus on the
solutions of the shape-invariance condition involving
translations of the parameters as shown in Eq. (\ref{10}). For this class 
the operator $\hat{T}(a_1)$ of
Eq. (\ref{9}) is simply given by
\begin{eqnarray}
  \label{12}
  \hat{T}(a_1) &=& \exp \left( \eta \frac{\partial}{\partial a_1}
  \right) \nonumber \\
  \hat{T}^{-1}(a_1) = \hat{T}^{\dagger}(a_1) &=& \exp \left( - 
\eta \frac{\partial}{\partial a_1} \right). 
\end{eqnarray}

Shape-invariant potentials are amenable to the treatment by the method
of creation and annihilation operators originally developed for the
harmonic oscillator. As such shape-invariant potentials are
generalizations of the harmonic oscillator potential. One must,
however, identify the creation and annihilation operators. The
obvious choice of $\hat{A}^{\dagger}$ and $\hat{A}$ does not work as
their commutator
\begin{equation}
  \label{12a}
  [\hat{A}, \hat{A}^{\dagger}]  = 2  \frac{\hbar}{\sqrt{2m}}
  \frac{dW}{dx}, 
\end{equation}
depends on the position. 
To establish the algebraic structure we first introduce the operators 
\begin{eqnarray}
  \label{13}
  \hat{B}_+ &=& \hat{A}^{\dagger}(a_1) \hat{T} (a_1) \nonumber \\
 \hat{B}_- = \hat{B}_+^{\dagger} &=& \hat{T}^{\dagger}(a_1) \hat{A} (a_1), 
\end{eqnarray}
and rewrite the Hamiltonian of Eq. (\ref{3}) as 
\begin{equation}
  \label{14}
  \hat{H} - E_0 = \hat{A}^{\dagger}\hat{A} = \hat{B}_+ \hat{B}_- .
\end{equation}
Using Eq. (\ref{8}) one can easily prove the commutation relation: 
\begin{equation}
  \label{15}
  [ \hat{B}_- , \hat{B}_+ ] = R(a_0) ,
\end{equation}
where we defined 
\begin{equation}
  \label{15a}
  a_n = a_1 + (n-1) \eta, 
\end{equation}
and used the identity 
\begin{equation}
  \label{16}
  R(a_n) = \hat{T}(a_1) R(a_{n-1})\hat{T}^{\dagger}(a_1),    
\end{equation}
valid for any $n$. 
Eq. (\ref{15}) suggests that $\hat{B}_-$ and $\hat{B}_+$ are the
appropriate creation and annihilation operators provided that their
non-commutativity with $R(a_1)$ is taken into account. Indeed 
using the relations 
\begin{eqnarray}
  \label{17}
   R(a_n) \hat{B}_+ &=& \hat{B}_+ R(a_{n-1}) \nonumber \\
   R(a_n) \hat{B}_- &=& \hat{B}_+ R(a_{n+1}), 
\end{eqnarray}
which readily follow from Eqs. (\ref{13}) and (\ref{16}), one can
write down the additional commutation relations 
\begin{equation}
  \label{18a}
  [\hat{H}, \hat{B}_+^n ] = (R(a_1)+R(a_2)+ \cdot \cdot + R(a_n))
  \hat{B}_+^n 
\end{equation}
and 
\begin{equation}
  \label{18b}
  [\hat{H}, \hat{B}_-^n ] = - \hat{B}_-^n (R(a_1)+R(a_2)+ \cdot \cdot
  + R(a_n))\,.
\end{equation}
Eqs. (\ref{18a}) and (\ref{18b}) are the generalization of the 
corresponding commutators
for the harmonic oscillator creation and annihilation
operators. Consequently the operators $\hat{B}_+$ and $\hat{B}_-$ can
be utilized as ladder operators for the spectra of the shape-invariant
potentials. Using Eqs. (\ref{4}) and (\ref{13}) one can show that the
ground state satisfies the condition
\begin{equation}
  \label{18c}
  \hat{B}_- | \psi_0 \rangle = 0.
\end{equation}
In the following we set the energy scale so that the ground state
energy, $E_0$, is zero. Using Eqs. (\ref{18a}) and (\ref{18c}) it
follows that
\begin{equation}
  \label{18d}
  \hat{H} \left( \hat{B}_+^n |\psi_0 \rangle \right) = (R(a_1)+R(a_2)+
  \cdot 
\cdot + R(a_n)) \left( \hat{B}_+^n |\psi_0 \rangle \right), 
\end{equation}
i.e., $\hat{B}_+^n |\psi_0 \rangle$ is an eigenstate of the
Hamiltonian 
with
the eigenvalue $R(a_1)+R(a_2)+ \cdot \cdot + R(a_n)$. Normalization
should be carried out with some care as $\hat{B}_+^n$ in general 
does not commute
with $R(a_n)$. One can show that the normalized wavefunction is 
\begin{equation}
  \label{18e}
  |\psi_n \rangle = \frac{1}{\sqrt{R(a_1)+ \cdot \cdot + 
R(a_n)}} \hat{B}_+ \cdot \cdot \frac{1}{\sqrt{R(a_1)+ 
R(a_2)}} \hat{B}_+  \frac{1}{\sqrt{R(a_1)}}\hat{B}_+
| \psi_0 \rangle. 
\end{equation}

In addition to the oscillator like commutation relations of
Eqs. (\ref{18a}) and (\ref{18b}) one gets new commutation relations
\begin{equation}
  \label{19}
  [ \hat{B}_+ , R(a_0) ] = (R(a_1) - R(a_0)) \hat{B}_+,  
\end{equation}
\begin{equation}
  \label{20}
  [ \hat{B}_+ , (R(a_1)-R(a_0))\hat{B}_+ ] = \{ (R(a_2) -
  R(a_1))-(R(a_1) - R(a_0))\} \hat{B}_+, 
\end{equation}
and so on. In general there is an infinite number of these commutation
relations. These commutation relations and their complex conjugates
along with Eq. (\ref{15}) form an infinite-dimensional Lie algebra,
realized here in a unitary representation. 

To classify algebras associated with the shape-invariant
potentials one can utilize the fact that for confining potentials 
the $n$th eigenvalue $E_n$
for large $n$ obeys the constraint \cite{nieto}
\begin{equation}
  \label{21}
  E_n \le {\rm constant} \times n^2. 
\end{equation}
Here we will show that those potentials where  $E_n$
is given by  
\begin{equation}
  \label{22}
  E_n = \beta n^2 + \delta n + \gamma, 
\end{equation}
lead to a finite shape-invariance algebra. Using Eq. (\ref{18d}) one 
can then show that 
\begin{equation}
  \label{23}
  R(a_n) = E_n - E_{n-1} = 2 \beta n + \delta - \beta, 
\end{equation}
which gives
\begin{equation}
  \label{24}
  R(a_n) - R(a_{n-1}) = 2 \beta.  
\end{equation}
Consequently for systems that satisfy Eq. (\ref{22}) the
resulting Lie algebra is finite:
\begin{equation}
  \label{25}
   [ \hat{B}_- , \hat{B}_+ ] = R(a_0), \,\,\,
 [ \hat{B}_+ , R(a_0) ] = 2 \beta \hat{B}_+.
\end{equation}
If $\beta$ is nonzero, depending on its sign, this algebra is either
$SU(2)$ or $SU(1,1)$. If $\beta$ is zero it is the Heisenberg-Weyl
algebra. For those potentials the eigenvalues of which which do not
satisfy Eq. (\ref{22}) the shape-invariance algebras remain 
infinite-dimensional. 
 

\section{Examples}

\indent

To illustrate the discussion in the previous section we first explicitly
work out two cases where one gets a finite Lie algebra and then give
the most general conditions on the superpotentials for the
shape-invariance Lie algebras to be finite. 

\subsection{Morse Potential}

\indent

For the Morse potential, $V(x) = V_0 ( e^{-2\lambda x} - 2 b e^{-\lambda
  x})$, the superpotential is 
\begin{equation}
  \label{26}
  W(x; a_n) = \sqrt{V_0} ( a_n - e^{- \lambda x}).
\end{equation}
The remainder in Eq. (\ref{7}) is given by
\begin{equation}         
  \label{27}
  R(a_n) = 2 \lambda \hbar \sqrt{\frac{V_0}{2m}} \left( a_n -
    \frac{1}{2} \frac{\lambda\hbar}{\sqrt{2mV_0}} \right), 
\end{equation}
where
\begin{equation}
  \label{28}
  a_n = b -  \frac{\lambda\hbar}{\sqrt{2mV_0}} (n-\frac{1}{2}). 
\end{equation}

Introducing the dimensionless operators 
\begin{equation}
  \label{29}
  K_0 = \frac{m}{\hbar^2\lambda^2} R(a_0)
\end{equation}
and 
\begin{equation}
  \label{30}
  K_{\pm} = \frac{\sqrt{m}}{\hbar\lambda} {\hat B}_{\pm},
\end{equation}
one finds that the shape-invariance algebra for the Morse potential is
$SU(1,1)$: 
\begin{equation}
  \label{31}
  [ K_+,K_-]=-2K_0, \>\> [K_0, K_{\pm}] = \pm K_{\pm}. 
\end{equation}
We note that this algebra is distinct from the $SU(2)$ algebra used in
the usual group-theoretical approach to the problem of 
finding bound-state solutions
of the one-dimensional Morse potential \cite{feza,levine}. As
\begin{displaymath}
  K_0 = \frac{\sqrt{2mV_0}}{\hbar \lambda}
\end{displaymath}
the shape-invariance algebra relates a series of Morse potentials with
different depths.

\subsection{Scarf Potential}

\indent

For the Scarf potential, $V(x) =-V_0/ \cosh^2 \lambda x$, the 
superpotential is
\begin{equation}
  \label{32}
   W(x; a_n) = \frac{\hbar \lambda}{\sqrt{2m}} a_n \tanh \lambda x,
\end{equation}
with 
\begin{equation}
  \label{33}
  R(a_n) = \frac{\hbar^2 \lambda^2}{2m} \left( 2a_n - 1  \right)
\end{equation}
where
\begin{equation}
  \label{34}
  a_n = \frac{1}{2} \left( \sqrt{\frac{8mV_0}{\hbar^2\lambda^2}+1}
    -2n+1 \right). 
\end{equation}
Again introducing the dimensionless operators
\begin{equation}
  \label{35}
  K_{\pm} = \frac{\sqrt{2m}}{\hbar \lambda} {\hat B}_{\pm}, 
\end{equation}
and 
\begin{equation}
  \label{36}
  K_0 =  \frac{m}{\hbar^2 \lambda^2} R(a_0)
\end{equation}
we again obtain the shape-invariance algebra to be an $SU(1,1)$
algebra: 
 \begin{equation}
  [ K_+,K_-]=-2K_0, \>\> [K_0, K_{\pm}] = \pm K_{\pm}. \nonumber  
\end{equation}

Once again the shape-invariance algebra relates a series of potentials
with different depths as $K_0$ is given as
\begin{equation}
  \label{37}
  K_0 = \left( \frac{2mV_0}{\hbar^2\lambda^2} + \frac{1}{4}
  \right)^{1/2}. 
\end{equation}

\subsection{General conditions for finite shape-invariance algebras}

\indent

Eqs. (\ref{15a}) and (\ref{23}) imply that when the shape-invariance 
algebra is finite $R(a_n)$ is linear in $a_n$, which in turn requires
the superpotential to be of the form 
\begin{equation}
  \label{38}
  W(x;a_n) = f(x) a_n + g(x). 
\end{equation}
For $R(a_n)$ to be independent of $x$, the functions $f(x)$ and $g(x)$
must satisfy the equations
\begin{equation}
  \label{39}
  \frac{\eta\hbar}{\sqrt{2m}} \frac{df}{dx} - \eta^2 f^2 = \beta,
\end{equation}
and
\begin{equation}
  \label{40}
  \frac{\hbar}{\sqrt{2m}} \frac{dg}{dx} -\eta f(x) g(x) =
  \frac{\delta}{2} + \frac{\beta a_1}{\eta}.
\end{equation}
The resulting $R(a_n)$ is of the form 
\begin{equation}
  \label{41}
   R(a_n) = 2 \beta a_n - \frac{\beta}{\eta} a_1 - \beta + \delta.
\end{equation}

One can then catalog those potentials for which the shape-invariance
algebra is finite-dimensional.  
Eq. (\ref{39}) indicates that $f(x)$ is a superpotential yielding to a
reflectionless potential. Thus it is either constant or proportional
to $\tanh x$. Eq. (\ref{40}) then implies that $g(x)$ is either
constant, proportional to $x$ or $\exp(x)$. This indicates that
constant potential, harmonic oscillator potential, Morse potential, 
and Scarf potential provide the complete list of potentials for which
the shape-invariance algebra is finite-dimensional. 

\section{Conclusions}

\indent

Shape invariance is shown to have 
an underlying algebraic structure and the associated Lie algebras are
identified. These Lie algebras transform the parameters of the
potentials such as strength or range. In general shape-invariance
algebras are infinite-dimensional. The conditions under which they
become finite-dimensional are elaborated. 

Shape-invariance was originally introduced in the context of one
dimensional quantum mechanics via the definition given in
Eq. (\ref{7}). However, it is possible to define shape-invariance in
terms of operators only, as given in Eq. (\ref{8}), without any
explicit reference to a potential function. An alternative approach 
to the study of 
quantum systems is to introduce algebraic Hamiltonians and exploit
dynamical symmetries associated with such Hamiltonians \cite{fr}. The
method introduced in this article can easily be extended to explore
shape-invariance properties of algebraic Hamiltonians. 
One such application is to utilize shape-invariance 
algebras in many-body problems where pairing plays an important
role. Indeed solutions for the generalized pairing Hamiltonian for 
spherical nuclei has been derived by introducing an
infinite-dimensional algebra \cite{lsu}. It may be possible to extend
this result to deformed nuclei using the techniques described here. 

It has been shown that harmonic oscillators with spin-orbit couplings
naturally lead to the superalgebras both as dynamical symmetry
algebras and spectrum-generating algebras \cite{baha1,baha2}. It would
be interesting to see if those results can be generalized to all
shape-invariant potentials. 

Finally, in this paper we omitted those shape-invariant potentials
where parameters are related by scaling. One expects that
shape-invariance algebras for such potentials can be related to
q-algebras. A detailed study of this connection is deferred to a later
publication. 


\section*{ACKNOWLEDGMENTS}

This work was
supported in part by the U.S. National Science Foundation Grant No.\
PHY-9605140 at the University of Wisconsin, and in part by the
University of Wisconsin Research Committee with funds granted by the
Wisconsin Alumni Research Foundation. We thank the 
Institute for Nuclear Theory and Department of Astronomy 
at the University of Washington for their hospitality and Department
of Energy for partial support during the completion of this work. 



\end{document}